\documentclass[reprint, nofootnotebib, nobibnotes, aip, jcp]{revtex4-2}
\setcitestyle{super}
\usepackage{graphicx}
\usepackage{amsmath}
\usepackage[bb=boondox]{mathalfa} %
\usepackage{bm}
\usepackage{physics}
\usepackage[version=3]{mhchem} %
\usepackage{comment}
\usepackage{soul}
\usepackage[LGR,T1]{fontenc}
\usepackage{tabularx}
\usepackage[normalem]{ulem} %
\usepackage{siunitx}
\DeclareSIUnit{\calorie}{cal}
\DeclareSIUnit{\Calorie}{\kilo\calorie}
\DeclareSIUnit{\angstrom}{\text {Å}}
\DeclareSIUnit{\electronmass}{\textit{m}_e}
\DeclareSIUnit{\hartree}{\textit{E}_{h}}
\DeclareSIUnit{\dalton}{u}

\usepackage{booktabs}
\usepackage{placeins} %

\usepackage{hyperref}
\hypersetup{colorlinks=true,linkcolor=blue,citecolor=blue,urlcolor=blue}
\usepackage{xcolor}
\newcommand{\red}[1]{{\leavevmode\color{red}{#1}}{}} %
\renewcommand{\red}[1]{{\leavevmode\color{black}{#1}}{}} %

\newcommand{\eu}{\mathrm{e}^}
\newcommand{\iu}{\ensuremath{\mathrm{i}}}
\newcommand{\rmd}{\mathrm{d}}
\newcommand{\half}{{\ensuremath{\frac{1}{2}}}}

\providecommand{\mat}[1]{\mathsf{#1}}
\renewcommand{\mathbf}[1]{\bm{#1}}
\renewcommand{\vec}[1]{\mathbf{\mathrm{#1}}}
\newcommand{\T}{\suptxt{T}}

\newcommand{\eqn}[1]{Eq.\,(\ref{#1})}
\newcommand{\eqns}[1]{Eqs.\,(\ref{#1})}

\newcommand{\Fig}[1]{Fig.~\ref{#1}}
\newcommand{\Figs}[1]{Figs.~\ref{#1}}

\newcommand{\Sec}[1]{Sec.~\ref{#1}}

\newcommand{\Refx}[1]{Ref.~\citenum{#1}}

\newcommand*\subtxt[1]{_{\textnormal{#1}}} %
\newcommand*\suptxt[1]{^{\textnormal{#1}}} %

\newcommand{\VR}{V\subtxt{R}}
\newcommand{\VB}{V\subtxt{B}}
\newcommand{\VP}{V\subtxt{P}}
\newcommand{\VRP}{V\subtxt{RP}}
\newcommand{\xRP}{\vec{x}\subtxt{RP}}
\newcommand{\VRB}{V\subtxt{RB}}
\newcommand{\VBP}{V\subtxt{BP}}
\newcommand{\VTS}{V^{\ddagger}}
\newcommand{\ER}{E\subtxt{R}^{\mu}}
\newcommand{\EP}{E\subtxt{P}^{\nu}}
\newcommand{\EB}{E\subtxt{B}^{\lambda}}
\newcommand{\eR}{\varepsilon\subtxt{R}}
\newcommand{\eB}{\varepsilon\subtxt{B}}
\newcommand{\eP}{\varepsilon\subtxt{P}}
\newcommand{\CR}{C\subtxt{R}}
\newcommand{\CB}{C\subtxt{B}}
\newcommand{\CP}{C\subtxt{P}}
\newcommand{\cff}{c}
\newcommand{\SR}{S\subtxt{R}}
\newcommand{\SB}{S_{\mathrm{B}}}
\newcommand{\SP}{S\subtxt{P}}
\newcommand{\KR}{\mathcal{K}\subtxt{R}}
\newcommand{\KB}{\mathcal{K}\subtxt{B}}
\newcommand{\KP}{\mathcal{K}\subtxt{P}}
\newcommand{\ZR}{Z\subtxt{R}}

\newcommand{\DeltaRB}{\Delta\subtxt{RB}}
\newcommand{\DeltaBP}{\Delta\subtxt{BP}}

\DeclareMathOperator{\diag}{diag}

\newcommand{\pder}[3][]{\frac{\partial^{#1}{#2}}{\partial{#3}^{#1}}}

\newcount\colveccount
\newcommand*\colvec[1]{
        \global\colveccount#1
        \begin{pmatrix}
        \colvecnext
}
\def\colvecnext#1{
        #1
        \global\advance\colveccount-1
        \ifnum\colveccount>0
                \quad
                \expandafter\colvecnext
        \else
                \end{pmatrix}^{\mathrm{T}}
        \fi
}
\newcount\rowveccount
\newcommand*\rowvec[1]{
        \global\rowveccount#1
        \begin{pmatrix}
        \rowvecnext
}
\def\rowvecnext#1{
        #1
        \global\advance\rowveccount-1
        \ifnum\rowveccount>0
                \ , \ 
                \expandafter\rowvecnext
        \else
                \end{pmatrix}
        \fi
}

\newcommand{\PaperTitle}{Uniform instanton theory for bridge-mediated nonadiabatic tunneling}

\begin{document}

\author{Imaad M. Ansari}
\altaffiliation{Present address: Centre for Computational Chemistry, School of Chemistry, University of Bristol, Bristol BS8 1TS, United Kingdom}
\affiliation{\mbox{Institute of Molecular Physical Sciences, ETH Z\"urich, 8093 Z\"urich, Switzerland}}

\author{George Trenins}
\altaffiliation{Present address: MPI for the Structure and Dynamics of Matter, Luruper Chaussee 149, 22761 Hamburg, Germany}
\affiliation{\mbox{Institute of Molecular Physical Sciences, ETH Z\"urich, 8093 Z\"urich, Switzerland}}

\author{Jeremy O. Richardson}
\email{jeremy.richardson@phys.chem.ethz.ch}
\affiliation{\mbox{Institute of Molecular Physical Sciences, ETH Z\"urich, 8093 Z\"urich, Switzerland}}

\date{\today}

\title{\PaperTitle}

\begin{abstract}
    Instanton theory provides a semiclassical approximation to nonadiabatic rate constants within a path-integral framework and captures nuclear quantum effects such as tunneling and zero-point energy.
    Although the simplest reactions proceed directly from reactants to products,
    a number of important nonadiabatic processes involve three electronic states and react via a concerted mechanism through a virtual bridge state, also referred to as bridge-mediated tunneling or superexchange.
    In previous work, we have derived an instanton theory to describe such reactions using a dominant tunneling trajectory that propagates on each of the three potential energy surfaces. %
    However, this approach is only valid when the bridge state is sufficiently low in energy.
    In contrast, in the high-bridge regime, the trajectories on the bridge surface become infinitesimally short, which causes the previous theory to break down. %
    In this work, we derive the appropriate asymptotic approximation for the high-bridge regime and find that the resulting theory reduces to a two-state golden-rule instanton with an effective coupling that describes a nonresonant hop to the virtual bridge state.
    We further develop a uniform asymptotic instanton theory that eliminates divergences and smoothly connects the high- and low-bridge regimes.
    In a companion paper, we apply the method to study the quintet-to-singlet transition (via a triplet bridge) of an iron(II) complex.
\end{abstract}

\maketitle

\section{Introduction}

The chemistry of molecular systems beyond the Born--Oppenheimer approximation is an area of great significance to numerous disciplines such as molecular electronics,\cite{SCO_application_1,SCO_memory_devices,molecular_switches_electronics,photomagnetization} organic photovoltaics,\cite{TADF_review} and quantum biology.\cite{quantumbiology_review_1,quantumbiology_review_3} 
In the nonadiabatic limit, the couplings between potential energy surfaces (PESs) in the diabatic representation are small, which allows for the use of perturbative methods to calculate nonadiabatic rate constants.

In the simplest case, where the initial reactant state decays directly into the product state, one can use Fermi's golden rule (FGR).\cite{FermiBook,DiracBook}
However, when the reactant and product states are not directly coupled to each other but only indirectly via a third bridge state, 
the rate constant is given in terms of second-order transitions via virtual bridge states and is
often referred to as ``superexchange.''\cite{KuehnBook,Jortner2002DNASuperex}
Examples include long-range, through-bond electron and energy transfer,\cite{Franzen1993SupexET,quantumbiology_review_1,voityuk_longrange_ET,gray_longrange_ET,bai_dexter,skourtis2016dexter} such as those in photosynthesis,\cite{Bixon1989_PhotosynthesysSuperexchange,Jortner1995BacRCSuperexET} double intersystem crossings,\cite{liesst_review_2,Hauser2004,Feterpy2_rocking,lundberg_transitionmetal_dimers,JortnerSecondOrderSOC} and Raman scattering.\cite{MukamelBook,Nitzan} 
However, in each case, the usual quantum-mechanical expression for the rate in terms of wavefunction overlaps is computationally intractable 
for complex molecules without making simplifying assumptions such as global harmonic approximations or dimensionality reduction.

Instanton theory provides an alternative framework to calculate rate constants of chemical reactions in full dimensionality including tunneling effects.\cite{Miller1975semiclassical,Perspective,InstReview} %
It is a semiclassical method that uses the path-integral representation of quantum mechanics\cite{Feynman} to describe the reaction in terms of an imaginary-time classical trajectory, called the instanton.
In addition to providing a reliable estimate for the rate,
the method gives mechanistic insight via the instanton trajectory itself, which can be interpreted as the dominant tunneling pathway.

Instanton theory has been generalized to nonadiabatic processes in the golden-rule limit,\cite{GoldenGreens,PhilTransA,inverted,weird,GRperspective}
for which
the tunneling pathway is determined by both the reactant and product PESs, and has been successfully applied to study tunneling in a number of intersystem-crossing and electron-transfer processes.\cite{thiophosgene,nitrene,CIinst,oxygen,carbene,housane}
Recently, we extended nonadiabatic instanton theory to bridge-mediated processes, in which part of the instanton trajectory travels on the bridge with an energy below the PES and therefore represents the mechanism for deep tunneling.\cite{PhilTransA} 
However, this bridge-mediated instanton theory is only applicable %
when the bridge drops below the reactant/product crossing seam.
As we shall show, for high bridges,
the asymptotic approximations involved in the derivation of our previous theory are no longer valid.

In this work, we extend bridge-mediated instanton theory to the high-bridge regime.
We begin by reviewing the low-bridge instanton derivation in \Sec{sec:low_bridge} and analyzing how the asymptotic approximations break down as the bridge is raised.
In \Sec{sec:high_bridge}, we derive a new instanton theory for the high-bridge regime.
In \Sec{sec:bridge_uniform}, we address the crossover region where the system transitions from the low- to the high-bridge regime and develop a uniform instanton approximation that is valid across both regimes.
Results are shown for a model system in \Sec{sec:results} as a proof of principle and conclusions are made in \Sec{sec:conclusions}.

The high-bridge regime is relevant for double intersystem crossings of transition-metal complexes.
In an accompanying paper,\cite{liesst} we apply the new methods with on-the-fly density functional theory in full dimensionality to study the quintet $\to$ triplet $\to$ singlet intersystem crossing following light-induced excited spin-state trapping (LIESST) in an iron(II) complex.\cite{Fepic3_liesst_OG,Hauser2004,Fepic3_PESs}

\section{Theory}

For a bridge-mediated process, 
the total Hamiltonian in the basis of the electronic states of the reactant $\ket{\mathrm{R}}$, bridge $\ket{\mathrm{B}}$ and product $\ket{\mathrm{P}}$ is\cite{KuehnBook} 
\begin{align} \label{eq:total_hamiltonian}
\hat{\mathcal{H}} = \sum_n  \hat{H}_n \ketbra{n}{n}  %
&+ \hat{\Delta}\subtxt{RB} \ketbra{\mathrm{R}}{\mathrm{B}} + \hat{\Delta}\subtxt{RB}^{\dagger} \ketbra{\mathrm{B}}{\mathrm{R}} \nonumber \\
& + \hat{\Delta}\subtxt{BP} \ketbra{\mathrm{B}}{\mathrm{P}} + \hat{\Delta}\subtxt{BP}^{\dagger}\ketbra{\mathrm{P}}{\mathrm{B}} ,
\end{align}
where $n \in \{ \mathrm{R, B, P} \}$, $\hat{\Delta}\subtxt{RB}$ and $\hat{\Delta}\subtxt{BP}$ are the reactant--bridge and  bridge--product diabatic couplings, and there is no direct coupling between the reactant and the product.  The nuclear Hamiltonians for each electronic state are expressed in terms of $f$-dimensional momentum ($\vec{p}$) and position ($\vec{x}$) vectors,
\begin{align} \label{eq:hamiltonian}
    \hat{H}_n = \frac{||\hat{\vec{p}}||^2}{2m} + V_n(\hat{\vec{x}}),
\end{align}
where $V_n(\vec{x})$ is the PES of state $n$.
Note that we use a mass-weighted coordinate system such that all nuclear degrees of freedom are associated with the same effective mass $m$.

\begin{figure}
	\centering
	\includegraphics{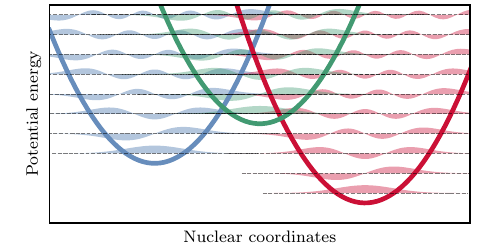}
	\caption{Potential energy surfaces and their vibrational eigenstates for a system undergoing a bridge-mediated reaction from the reactant (blue) to the product (red) via the bridge (green). }
	\label{fig:bridge_overlaps}
\end{figure}

The vibrational eigenstates of the reactant, bridge and product Hamiltonians %
obey the time-independent Schr\"odinger equations
\begin{subequations} \label{eq:eigenvalue_hamiltonian}
\begin{align} 
    \hat{H}\subtxt{R} \ket{\chi\subtxt{R}^{\mu}} &= \ER  \ket{\chi\subtxt{R}^{\mu}} , \\
    \hat{H}\subtxt{B} \ket{\chi\subtxt{B}^{\lambda}} &= \EB \ket{\chi\subtxt{B}^{\lambda}}  , \\
    \hat{H}\subtxt{P} \ket{\chi\subtxt{P}^{\nu}} &= \EP \ket{\chi\subtxt{P}^{\nu}} .
\end{align}
\end{subequations}
An illustration of the PESs and their vibrational wavefunctions %
are shown in \Fig{fig:bridge_overlaps}.

The quantum-mechanical rate constant for a bridge-mediated reaction from an initial thermal ensemble of reactant states to the product states can be written using second-order time-dependent perturbation theory as\cite{DiracBook,Feynman,KuehnBook,PhilTransA}
\begin{align} \label{eq:rate_bridge}
    k \ZR  &= \frac{2\pi}{\hbar} \sum_{\mu} \sum_{\nu} \eu{-\beta \ER} \, | \mathcal{T}_{\mu \nu} |^2 \, \delta(\ER - \EP) , %
\end{align}
where $\ZR=\sum_\mu\eu{-\beta\ER}$ is the reactant partition function, $\beta = 1/k\subtxt{B}T$ is the inverse temperature, and
\begin{equation} \label{eq:transition_amplitude}
    \mathcal{T}_{\mu \nu} = \sum_{\lambda} \frac{\bra{\chi\subtxt{R}^{\mu}} \hat{\Delta}\subtxt{RB}  \ket{\chi\subtxt{B}^{\lambda}} \bra{\chi\subtxt{B}^{\lambda}} \hat{\Delta}\subtxt{BP} \ket{\chi\subtxt{P}^{\nu}}}{\EB - \ER} 
\end{equation} 
is the amplitude of a second-order transition from %
state $\mu$ of the reactant to state $\nu$ of the product
via the virtual bridge states $\lambda$.

However, using \eqn{eq:rate_bridge} as written requires solving the time-independent Schr\"odinger equations [\eqn{eq:eigenvalue_hamiltonian}], which is computationally infeasible except for simple analytic or low-dimensional Hamiltonians.
In this work, we will derive semiclassical instanton approximations to \eqn{eq:rate_bridge} in the high- and low-bridge regimes.
The instanton methods are far more computationally efficient and can thus be applied to real molecular systems in full dimensionality, as we have demonstrated in other work.\cite{thiophosgene,nitrene,CIinst,oxygen,carbene,housane,liesst}

\subsection{Low-bridge regime} \label{sec:low_bridge}

We start with a review of bridge-mediated instanton theory in the low-bridge regime, which was first derived in \Refx{PhilTransA}, as it provides much of the necessary formalism for the extension to the high-bridge regime in \Sec{sec:high_bridge}.\@
First, \eqn{eq:rate_bridge} is reformulated in terms of a three-time correlation function \cite{PhilTransA}\textsuperscript{,}\footnote{Note that this is similar but not identical to the three-time correlation functions derived in \Refx{KuehnBook}}
\begin{align} \label{eq:k4_cff}
    k \ZR = \frac{1}{\hbar^4} \int_{-\infty}^{\infty}  \rmd t  \iint_{0}^{\infty} \rmd \tau' \rmd \tau'' \, \cff (z, \tau', \tau'') ,
\end{align}
where we treat time as a complex variable, $z = \tau + \iu t$, and are free to choose the imaginary-time shift $\tau$.\footnote{One can choose $\tau$ arbitrarily within the holomorphic region of the correlation function.  This covers at least the range $[0,\beta\hbar]$, but can often be extended to negative values to cover the inverted regime, as shown in \Refx{inverted}, until a branch point is encountered, as explained in \Refx{oxygen}.}
In the path-integral formalism, the correlation function,
\begin{align} \label{eq:cff}
    \cff(z, \tau', \tau'') &=  \iiiint_{-\infty}^{+\infty}  \rmd \vec{x}\subtxt{R}' \, \rmd\vec{x}\subtxt{P}' \, \rmd\vec{x}\subtxt{R}'' \, \rmd\vec{x}\subtxt{P}'' \nonumber \\
    & \quad \times \DeltaRB^{*}(\vec{x}\subtxt{R}'') \, \Delta\subtxt{RB}(\vec{x}\subtxt{R}') \, \Delta\subtxt{BP}(\vec{x}\subtxt{P}') \, \DeltaBP^{*}(\vec{x}\subtxt{P}'') \nonumber \\
	 &\quad \times  \KR(\vec{x}\subtxt{R}'', \vec{x}\subtxt{R}', \beta\hbar - z - \tau') \, \KB(\vec{x}\subtxt{R}', \vec{x}\subtxt{P}', \tau')  \nonumber \\
	 &\quad \times \KP(\vec{x}\subtxt{P}', \vec{x}\subtxt{P}'', z - \tau'') \, \KB(\vec{x}\subtxt{P}'', \vec{x}\subtxt{R}'', \tau''),
\end{align}
is expressed in terms of imaginary-time propagators $\mathcal{K}_n(\vec{x}_b, \vec{x}_a, \tau_n) = \bra{\vec{x}_b} \eu{-\hat{H}_n \tau_n/\hbar} \ket{\vec{x}_a}$, which give the quantum amplitude for the nuclei to travel from $\vec{x}_a$ to $\vec{x}_b$ in imaginary time $\tau_n$ according to the Hamiltonian $\hat{H}_n$.

The propagators are approximated by the semiclassical van-Vleck form \cite{Miller1971density,InstReview}
\begin{align} \label{eq:Kn}
    \mathcal{K}_n (\vec{x}_b, \vec{x}_a, \tau_n) \simeq \sqrt{\frac{C_n (\vec{x}_b, \vec{x}_a, \tau_n)}{(2\pi\hbar)^f}} \; \eu{-S_n (\vec{x}_b, \vec{x}_a, \tau_n)/\hbar} ,
\end{align}
where
\begin{equation}
    S_n(\vec{x}_b, \vec{x}_a , \tau_n) = \int_0^{\tau_n} \rmd \varsigma \, \Big[ \tfrac{1}{2} m||\dot{\vec{x}}(\varsigma)||^2 + V_n(\vec{x}(\varsigma)) \Big]
\end{equation}
is the Euclidean action of a classical trajectory from $\vec{x}_a$ to $\vec{x}_b$ in imaginary time $\tau_n$ and
\begin{align} \label{eq:Cn}
    C_n = \det \left[ - \pdv{S_n}{\vec{x}_a}{\vec{x}_b} \right]
\end{align}
accounts for the fluctuations about the classical trajectory.\cite{GutzwillerBook,Kleinert}

The instanton approximation in the low-bridge regime is obtained by a steepest-descent integration of \eqn{eq:k4_cff} with the semiclassical correlation function [\eqn{eq:cff} with the propagators from \eqn{eq:Kn}] over all variables ($\vec{x}\subtxt{R}'$, $\vec{x}\subtxt{P}'$, $\vec{x}\subtxt{P}''$, $\vec{x}\subtxt{R}''$, $z$, $\tau'$, $\tau''$) simultaneously.
The paths that contribute to this integral are composed of segments on the reactant, bridge and product states with the combined action
\begin{align}
    S^{(4)}&(\vec{x}\subtxt{R}', \vec{x}\subtxt{P}', \vec{x}\subtxt{P}'', \vec{x}\subtxt{R}'', z, \tau', \tau'') \nonumber \\
     & = \SR(\vec{x}\subtxt{R}'', \vec{x}\subtxt{R}', \beta\hbar - z - \tau') + \SB(\vec{x}\subtxt{R}', \vec{x}\subtxt{P}', \tau') \nonumber \\
	& \quad +  \SP(\vec{x}\subtxt{P}', \vec{x}\subtxt{P}'', z - \tau'') + \SB(\vec{x}\subtxt{P}'', \vec{x}\subtxt{R}'', \tau'') .
\end{align}
The dominant path is determined by a stationary point of the action with respect to all seven variables (denoted as $\tilde{\vec{x}}\subtxt{R}'$, $\tilde{\vec{x}}\subtxt{P}'$, $\tilde{\vec{x}}\subtxt{P}''$, $\tilde{\vec{x}}\subtxt{R}''$, $\tilde{\tau}$, $\tilde{\tau}'$, $\tilde{\tau}''$).
This stationary-action trajectory is called the instanton and it represents the optimal tunneling path.
Following the mathematical procedure of steepest-descent integration,\cite{BenderBook}
we perform a second-order Taylor expansion about the stationary point and integrate over the resulting Gaussian function, assuming slowly varying prefactors, to obtain the low-bridge (LB) instanton rate constant:\cite{PhilTransA}
\begin{align} \label{eq:k4_inst}
    k\subtxt{LB} \ZR = (2\pi\hbar)^{3/2} \frac{|\DeltaRB \DeltaBP|^2}{\hbar^4}  \sqrt{\frac{\CR \CB' \CP \CB''}{-\Sigma^{(4)}}} \; \eu{-S^{(4)}/\hbar} ,
\end{align}
where the fluctuation factors $\CR$ and $\CP$ are defined according to \eqn{eq:Cn}, and similarly 
\begin{subequations}
\begin{align}
    \CB' &= \det \left[ - \pdv{\SB(\vec{x}\subtxt{R}', \vec{x}\subtxt{P}', \tau')}{\vec{x}\subtxt{R}'}{\vec{x}\subtxt{P}'} \right] , \\
    \CB'' &= \det \left[ - \pdv{\SB(\vec{x}\subtxt{P}'', \vec{x}\subtxt{R}'', \tau'')}{\vec{x}\subtxt{P}''}{\vec{x}\subtxt{R}''} \right] .
\end{align}
\end{subequations}
Finally, $\Sigma^{(4)} = \det[\mat{H}]$ is the determinant of the Hessian of $S^{(4)}$ with respect to $\vec{x}\subtxt{R}'$, $\vec{x}\subtxt{P}'$, $\vec{x}\subtxt{P}''$, $\vec{x}\subtxt{R}''$, $\tau$, $\tau'$ and $\tau''$.
The negative sign before $\Sigma^{(4)}$ arises as a result of the Cauchy--Riemann identities,\cite{ComplexVariables} which express the derivatives of $S^{(4)}$ with respect to $t$ in terms of derivatives with respect to $\tau$.
All quantities in \eqn{eq:k4_inst} including the action, the diabatic couplings, the fluctuation factors and $\Sigma^{(4)}$ are evaluated at the stationary point.

\begin{figure*}
	\centering
	\includegraphics{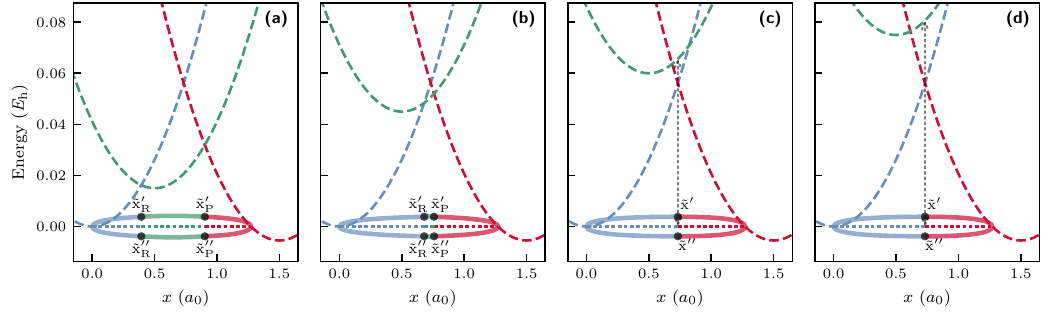}
	\caption{The low-temperature instanton for the model system in (a) the low-bridge regime [$\eB = \SI{0.015}{\hartree}$], (b) borderline low-bridge regime [$\eB = \SI{0.045}{\hartree}$], (c) borderline high-bridge regime [$\eB = \SI{0.060}{\hartree}$] and (d) the high-bridge regime [$\eB = \SI{0.075}{\hartree}$].
    The potential energy surfaces are plotted as dashed lines, and the instanton as a dotted line at the tunneling energy.
	Solid translucent lines are used to illustrate the forward and backward segments of the instanton with an arbitrary spacing along the $y$ axis.
	On these curves, the hopping points are depicted with black dots.
    The vertical arrows in (c) and (d) are used to depict the nonresonant hop to the infinitesimally short trajectory on $\VB$.
	The colors correspond to the reactant (blue), bridge (green) and product (red).
    }
	\label{fig:inst_bridge_regimes}
\end{figure*}

In order to illustrate the behavior of this bridge-mediated instanton theory, we consider a model system of three harmonic oscillators
\begin{subequations} \label{eq:model_system}
\begin{align}
   \VR(x) &= \tfrac{1}{2}  m\omega^2 x^2 + \eR, \\
   \VB(x) &= \tfrac{1}{2} m\omega^2 (x - x\subtxt{B})^2 + \eB , \\
   \VP(x) &=  \tfrac{1}{2}  m\omega^2 (x - x\subtxt{P})^2 + \eP,
\end{align}
\end{subequations}
with $m = 7000\,m\subtxt{e}$, $\omega = \SI{1200}{\per\cm}$, $x\subtxt{B} = \SI{0.5}{\bohr}$, $x\subtxt{P} = \SI{1.5}{\bohr}$, $\eR=0$, $\eP = -\hbar\omega$, and $\eB > 0$ is varied.
We use a one-dimensional harmonic model system, as its correlation function and the quantum-mechanical rate constant can be evaluated exactly.
However, it is important to note that the instanton formalism is also applicable to multidimensional anharmonic PESs, for which efficient numerical methods have been developed,\cite{InstReview,PhilTransA}
enabling its application to realistic molecular systems in full dimensionality.\cite{thiophosgene,nitrene,CIinst,oxygen,carbene,housane,liesst}

The instanton, shown in \Fig{fig:inst_bridge_regimes}(a), is a periodic orbit, with hops between $\VR$ and $\VB$ at $\tilde{\vec{x}}\subtxt{R}'$ and $\tilde{\vec{x}}\subtxt{R}''$, and between $\VP$ and $\VB$
at $\tilde{\vec{x}}\subtxt{P}'$ and $\tilde{\vec{x}}\subtxt{P}''$.
The stationary-action conditions ensure that the momentum is conserved at the hopping points and that the energies of all four trajectories are equal, which in turn constrains the hopping points to the corresponding crossing seams.\cite{GoldenGreens,PhilTransA}
The instanton energy is below all bridge states and thus represents the reaction mechanism for a single-step, concerted tunneling process via virtual bridge states.

Note that the bridge-mediated instanton formalism discussed in this work is only valid
at low temperature, where an instanton exists with an energy below the bridge.\cite{PhilTransA}
At high temperatures,
a two-step sequential mechanism would be expected to give dominant contributions to the rate constant.\cite{UlstrupBook}
Assuming that the system would thermalize after hopping to the bridge, the rate constant for the sequential mechanism could be easily calculated using two standard golden-rule instantons and applying the basic rules of chemical kinetics for a two-step reaction.  In the following, we focus on the more interesting concerted case.

However, as $\eB$ is increased, the imaginary time spent on the bridge ($\tilde{\tau}'$ and $\tilde{\tau}''$) becomes shorter [\Fig{fig:inst_bridge_regimes}(b)].
For a sufficiently large $\eB$, 
the bridge PES goes higher than the reactant--product crossing point, which changes the topology of the problem completely [\Fig{fig:inst_bridge_regimes}(c--d)].
In principle, a stationary-action instanton pathway still exists that hops at the RB and BP crossing points.
However, in this case, 
the stationary points $\tilde{\tau}'$ and $\tilde{\tau}''$ would have to be negative, as it would otherwise be impossible to satisfy both momentum and energy conservation (which are consequences of the stationary-point conditions with respect to positions and the time variables), analogous to the inverted regime for golden-rule instantons.\cite{inverted} 
As these stationary points $\tilde{\tau}'$ and $\tilde{\tau}''$ are outside the integration bounds of \eqn{eq:k4_cff},
the Gaussian steepest-descent integration employed to obtain \eqn{eq:k4_inst} is no longer valid and a new approach is required. %

We therefore identify two regimes based on the relative energies of $\VR$, $\VP$ and $\VB$ at the reactant--product %
crossing point, $\xRP$, with energy $\VRP$:
\begin{itemize}
	\item Low-bridge regime: $\VB(\xRP) < \VRP$;
	\item High-bridge regime: $\VB(\xRP) > \VRP$.
\end{itemize}
These regimes are illustrated in \Fig{fig:inst_bridge_regimes}.

The instanton approach outlined in this section %
is only valid in the low-bridge regime (as long as the temperature is low enough that the instanton exists).
In \Sec{sec:high_bridge}, we derive a new bridge-mediated instanton theory for the high-bridge regime. %
We will also 
develop a uniform asymptotic approximation in \Sec{sec:bridge_uniform} to
address the crossover region where the system transitions between the two regimes. %

\subsection{High-bridge regime} \label{sec:high_bridge}

\begin{figure}
	\centering
	\includegraphics{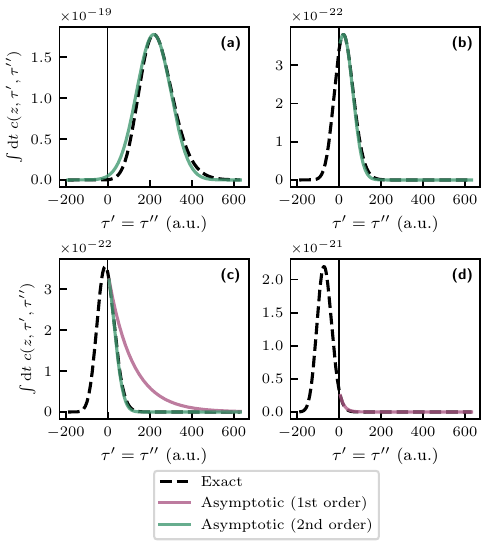}
	\caption{The two-time correlation function and its asymptotic approximation in  (a) the low-bridge regime [$\eB = \SI{0.015}{\hartree}$], (b) borderline low-bridge regime [$\eB = \SI{0.045}{\hartree}$], (c) borderline high-bridge regime [$\eB = \SI{0.060}{\hartree}$] and (d) the high-bridge regime [$\eB = \SI{0.075}{\hartree}$].
	The two-time correlation function is obtained by integrating $\cff(z, \tau', \tau'')$ along $t$ by the method of steepest descent about the stationary point in $z$, $\tilde{\tau}(\tau', \tau'')$. }
	\label{fig:bridge_cff}
\end{figure}

To fully understand the reasons for the breakdown of the approach of \Sec{sec:low_bridge} in the high-bridge regime,
we study the correlation functions shown in \Fig{fig:bridge_cff}.
In the low-bridge regime [\Fig{fig:bridge_cff}(a)], the stationary point (in this case the maximum) is found at a positive value of $\tau'=\tau''$ and the integral is well approximated by %
a Gaussian centred on the stationary point.
The first derivative at the top of a Gaussian is zero, and so the leading-order asymptotic approximation requires a second-order expansion.
However, in the high-bridge regime [\Fig{fig:bridge_cff}(d)],
the stationary point is located at a negative value of $\tau'=\tau''$, which is outside the integration bounds.
The dominant value of the integrand is now at the integration boundary, $\tau' = \tau'' = 0$, and 
the correct way\cite{BenderBook} to take the leading-order asymptotic approximation to the integral over $\tau'$ and $\tau''$ is to expand the terms in the integrand about this point. %
In this case, a first-order expansion is sufficient.

As we are performing the expansion about $\tilde{\tau}' = \tilde{\tau}'' = 0$, we can use the short-time limit\cite{Feynman} of the imaginary-time propagator in \eqn{eq:Kn} on the bridge, valid as $\tau'\to0$, 
\begin{align}
    S\subtxt{B}(\vec{x}\subtxt{R}', \vec{x}\subtxt{P}', \tau')
    &\simeq \frac{m||\vec{x}\subtxt{R}' - \vec{x}\subtxt{P}'||^2}{2\tau'} + \tau' \VB \left( \frac{\vec{x}\subtxt{R}' + \vec{x}\subtxt{P}'}{2} \right) \nonumber \\
    C\subtxt{B}(\vec{x}\subtxt{R}', \vec{x}\subtxt{P}', \tau')
    &\simeq \left( \frac{m}{\tau'} \right)^{f} ,
    \label{eq:shorttime}
\end{align}
and equivalently for the other bridge propagator with double primes. %
Therefore, together with the following variable transformation:
\begin{subequations} \label{eq:highbridge_transform}
\begin{align}
	\vec{x}' &= \frac{\vec{x}\subtxt{R}' + \vec{x}\subtxt{P}'}{2} , \\
	\vec{x}'' &= \frac{\vec{x}\subtxt{R}'' + \vec{x}\subtxt{P}''}{2} , \\
	\vec{x}_{-}' &= \vec{x}\subtxt{P}' - \vec{x}\subtxt{R}' , \\
	\vec{x}_{-}'' &= \vec{x}\subtxt{P}'' - \vec{x}\subtxt{R}'' ,
\end{align}
\end{subequations}
we can write the total action as
\begin{align}
	S^{(4)} \simeq & \; \SR \left( \vec{x}'' -  \frac{\vec{x}_{-}''}{2}, \vec{x}' - \frac{\vec{x}_{-}'}{2}, \beta\hbar - z - \tau' \right) \nonumber \\
    & +   \SP \left( \vec{x}' +  \frac{\vec{x}_{-}'}{2}, \vec{x}'' +  \frac{\vec{x}_{-}''}{2}, z - \tau'' \right) \nonumber \\
     & +  \frac{m}{2\tau'}||\vec{x}_{-}'||^2 + \tau' \VB(\vec{x}') + \frac{m}{2\tau''}||\vec{x}_{-}''||^2 + \tau'' \VB(\vec{x}'') .
\end{align}

The terms proportional to $1/\tau'$ and $1/\tau''$
cause difficulties for performing the standard steepest-descent procedure
as they cannot be represented by a Taylor series about $\tilde{\tau}'=\tilde{\tau}''=0$.
However, instead of performing the integrals over positions and time simultaneously, we will do them one after the other, which, in this case, resolves the problem.
As we shall show in the following, integrating over $\vec{x}_{-}'$ and $\vec{x}_-''$ eliminates the singular terms, leaving an effective action that can be expanded as a Taylor series up to first order in $\tau'$ and $\tau''$, which is easily integrated.

Using \eqn{eq:highbridge_transform} in Eqs.~\eqref{eq:k4_cff} and \eqref{eq:cff}, we
first integrate out $\vec{x}_-'$ and $\vec{x}_-''$ by steepest descent around their stationary points to give
\begin{multline} \label{eq:rate_highbridge_int1}
	k \ZR \simeq \int_{-\infty}^{+\infty} \rmd t \iint_{0}^{\infty}  \rmd \tau' \rmd \tau'' \iint_{-\infty}^{+\infty} \rmd \vec{x}' \rmd \vec{x}'' \\ 
    \times \frac{|\DeltaRB \DeltaBP|^2}{\hbar^4}   \frac{\sqrt{\CR \CP}}{(2\pi\hbar)^f} \sqrt{\frac{\CB'\CB''}{\det[\mat{A}]}} \; \eu{-\gamma/\hbar} .
\end{multline}
The new effective action is
\begin{multline} \label{eq:effective_action}
    \gamma(\vec{x}', \vec{x}'', z, \tau', \tau'')
    = S^{(4)}(\tilde{\vec{x}}_-',\tilde{\vec{x}}_-'',\vec{x}',\vec{x}'',z,\tau',\tau'') ,
\end{multline}
where $\tilde{\vec{x}}_-'$ and $\tilde{\vec{x}}_-''$ are the $(\vec{x}',\vec{x}'',z,\tau',\tau'')$-dependent solutions of $\pder{S^{(4)}}{\vec{x}_-'}=\vec{0}$ and $\pder{S^{(4)}}{\vec{x}_-''}=\vec{0}$.
In particular, using the relations between derivatives of the action and 
the imaginary-time momentum $\vec{p}_n$ of the trajectory %
on state $n$,\cite{LandauMechanics,GutzwillerBook,InstReview} we obtain
\begin{subequations}
\label{eq:pB}
\begin{align} 
    \frac{m\tilde{\vec{x}}_-'}{\tau'} &= -\half(\vec{p}\subtxt{R}' + \vec{p}\subtxt{P}') \equiv -\vec{p}\subtxt{B}', \\
    \frac{m\tilde{\vec{x}}_-''}{\tau''} &= \half(\vec{p}\subtxt{R}'' + \vec{p}\subtxt{P}'') \equiv \vec{p}\subtxt{B}''.
\end{align}
\end{subequations}
The fluctuation factor in \eqn{eq:rate_highbridge_int1} is determined by
\begin{align}
    \mat{A} &= \pdv{S^{(4)}}{\vec{x}_{-}}{\vec{x}_{-}}  \\ \nonumber
    & = \begin{pmatrix}
		\frac{m}{\tau'}\mathbb{I}_{f} & \mathbb{0}_f \\
		\mathbb{0}_f & \frac{m}{\tau''} \mathbb{I}_{f} 
	\end{pmatrix} + \frac{1}{  4} \begin{pmatrix}
		 \pdv{(\SR + \SP)}{\vec{x}'}{\vec{x}'} & \pdv{(\SR + \SP)}{\vec{x}'}{\vec{x}''} \\[0.5em]
		\pdv{(\SR + \SP)}{\vec{x}''}{\vec{x}'} & \pdv{(\SR + \SP)}{\vec{x}''}{\vec{x}''}
		\end{pmatrix} ,
\end{align}
where $\vec{x}_{-} =(\vec{x}_-',\vec{x}_-'')$,
$\mathbb{I}_f$ is the $f \times f$ identity matrix and $\mathbb{0}_f$ is an $f \times f$ matrix of zeros.
To obtain the leading-order asymptotic approximation, we only require the prefactor in the limit
$\tau', \tau'' \to 0$, for which
$\det[\mat{A}]\simeq C\subtxt{B}'C\subtxt{B}''$, leading to a cancellation in the rate expression \eqn{eq:rate_highbridge_int1}.

Next we integrate over $\vec{x}'$, $\vec{x}''$ and $t$ by steepest descent to give
\begin{align} \label{eq:kHB_tauint}
	k \ZR %
    &\simeq \sqrt{2\pi\hbar} \iint_{0}^{\infty}  \rmd \tau' \rmd \tau'' \frac{|\DeltaRB \DeltaBP|^2}{\hbar^4} \sqrt\frac{\CR \CP}{-\det[\Gamma]} \, \eu{-\phi/\hbar} ,
\end{align}
where %
\begin{align}
    \phi(\tau',\tau'') &= %
    S^{(4)}(\tilde{\vec{x}}_-',\tilde{\vec{x}}_-'',\tilde{\vec{x}}',\tilde{\vec{x}}'',\tilde{\tau},\tau',\tau'') ,
\end{align}
and $\tilde{\vec{x}}'$, $\tilde{\vec{x}}''$ and $\tilde{\tau}$ are the $(\tau',\tau'')$-dependent solutions of $\pder{\gamma}{\vec{x}'}=\vec{0}$, $\pder{\gamma}{\vec{x}''}=\vec{0}$ and $\pder{\gamma}{\tau}=0$.
The fluctuation matrix $\Gamma$ is the Hessian of $\gamma$ with respect to $\vec{x}'$, $\vec{x}''$ and $\tau$, evaluated at the stationary point.
Again, we only require this in the limit $\tau', \tau'' \to 0$, in which it reduces to %
the Hessian of $S^{(2)}$, called $\Sigma^{(2)}$, %
where
$S^{(2)}$ is the action used in the two-state, golden-rule instanton theory \cite{GoldenGreens,PhilTransA,GRperspective}
\begin{align}
    S^{(2)}(\vec{x}', \vec{x}'', \tau) &= \SR(\vec{x}'', \vec{x}', \beta\hbar - \tau) + \SP(\vec{x}', \vec{x}'', \tau) ,
\end{align}
with hopping points at $\vec{x}'$ and $\vec{x}''$.

As we explained above, in order to obtain the leading-order asymptotic approximation to the final integral, it is only necessary to expand $\phi$ to first order around $\tilde{\tau}'=\tilde{\tau}''=0$:
\begin{align} \label{eq:phiTaylor}
    \phi(\tau',\tau'') &\simeq S^{(2)} + \tau' \mathcal{E}' + \tau'' \mathcal{E}'',
\end{align}
with
\begin{subequations} \label{eq:mathcalE}
\begin{align}
    \mathcal{E}' &= \pder{S^{(4)}}{\tau'} = -E\subtxt{R} + V\subtxt{B}(\tilde{\vec{x}}') - \frac{||\vec{p}\subtxt{B}'||^2}{2m} ,
    \\
    \mathcal{E}'' &= \pder{S^{(4)}}{\tau''} = -E\subtxt{P} + V\subtxt{B}(\tilde{\vec{x}}'') - \frac{||\vec{p}\subtxt{B}''||^2}{2m} ,
\end{align}
\end{subequations}
where %
the energies of the respective instanton segments\cite{LandauMechanics,GutzwillerBook,InstReview} are $E\subtxt{R} = \pdv{\SR}{\tau\subtxt{R}}$ and $E\subtxt{P} = \pdv{\SP}{\tau\subtxt{P}}$,
with $\tau\subtxt{R}=\beta\hbar-\tilde{\tau}$ and $\tau\subtxt{P}=\tilde{\tau}$ as for the golden-rule instanton.\cite{GoldenGreens}

Finally, integrating over $\tau'$ and $\tau''$ gives the high-bridge (HB) instanton rate constant, 
\begin{align} \label{eq:kHB}
    k\subtxt{HB} \ZR = \sqrt{2\pi\hbar} \, \frac{|\Delta\subtxt{SEC}\suptxt{inst}|^2}{\hbar^2} \sqrt{\frac{\CR\CP}{-\Sigma^{(2)}}} \; \eu{-S^{(2)}/\hbar} ,
\end{align}
which is identical to the two-state golden-rule case\cite{GoldenGreens} except with
an effective coupling
\begin{align} \label{eq:Delta_highbridge_int1}
    |\Delta\subtxt{SEC}\suptxt{inst}| = \frac{|\DeltaRB \DeltaBP|}{\sqrt{\mathcal{E}' \mathcal{E}''}} .
\end{align}

At the stationary point of $S^{(2)}$, the instanton is a periodic orbit such that %
$\vec{p}\subtxt{B}' = \vec{p}\subtxt{R}' = \vec{p}\subtxt{P}'$ and $\vec{p}\subtxt{B}'' = \vec{p}\subtxt{R}'' = \vec{p}\subtxt{P}''$. 
Using the expression for the energy of an imaginary-time trajectory, $E_n = V_n - ||\vec{p}_n||^2/2m $, \eqn{eq:mathcalE} thus simplifies to potential-energy differences at the hopping points $\tilde{\vec{x}}'$ and $\tilde{\vec{x}}''$:
\begin{subequations} \label{eq:energydiff_MECP}
\begin{align}
    \mathcal{E}' &=  \VB(\tilde{\vec{x}}') - \VR(\tilde{\vec{x}}') ,  \\
    \mathcal{E}'' &= \VB(\tilde{\vec{x}}'') - \VP(\tilde{\vec{x}}'') .
\end{align}
\end{subequations}
The stationary-point conditions also ensure that the hopping points are located on the crossing seam,\cite{GoldenGreens} and as the instanton typically folds back on itself such that the two hopping points are coincident, we have $\VR(\tilde{\vec{x}}') = \VP(\tilde{\vec{x}}'')=\VRP^\ddagger$.
Therefore, the effective coupling can be formulated as
\begin{align} \label{eq:Delta_highbridge}
    |\Delta\subtxt{SEC}\suptxt{inst}| = \frac{|\DeltaRB \DeltaBP|}{\VB^{\ddagger} - \VRP^\ddagger} ,
\end{align}
where $\VB^{\ddagger}$ and $\VRP^\ddagger$ are the bridge and reactant/product potentials at the instanton hopping point.
Note that as %
the effective coupling [and hence the entire right-hand side of \eqn{eq:kHB}] is equivalent for the forwards and backwards reactions,
the overall rate theory obeys detailed balance.

In fact, golden-rule rate theory with an effective coupling is a commonly used approximation for bridge-mediated reactions and is also known as the superexchange effective coupling (SEC) approach.\cite{ratner1990bridge,KuehnBook} 
It is usually obtained by ignoring the vibrational energy-level structure %
in \eqn{eq:transition_amplitude} and instead using
\red{the pure electronic energy difference (EEG), %
\begin{align} \label{eq:Delta_highbridge_SEC} %
	|\Delta\subtxt{SEC}\suptxt{EEG}| = \frac{|\DeltaRB\DeltaBP|}{\varepsilon\subtxt{B}-\varepsilon\subtxt{R}} . 
\end{align}}%
\red{May and Kühn\cite{KuehnBook} point out that this approximation is only expected to be valid when the bridge levels are well separated from the reactant and product.
Nontheless, due to its simplicity, such approximations are commonly found in the literature.\cite{Bixon2002,ShimazakiDNASuperexchange,JortnerSecondOrderSOC,geng2015superexchange,Jortner2002DNASuperex}}
As we will show, the instanton approach is far more accurate than this approximation, as one would expect from its rigorous asymptotic derivation.
In addition, %
the approximation \eqn{eq:Delta_highbridge_SEC} breaks the detailed-balance condition that is obeyed by the instanton method.

Our effective coupling \eqn{eq:Delta_highbridge} is however the natural generalization of
another form present in the literature,
which is based on the potential difference at the minimum-energy crossing point (MECP). %
\cite{Bixon1989_PhotosynthesysSuperexchange}
In most one-dimensional systems, this is equivalent to the hopping point of the instanton and is thus identical to \eqn{eq:Delta_highbridge}.
However, in multidimensional systems, even though
the hopping point is usually constrained to lie on the crossing seam,\cite{inverted} corner-cutting effects may cause it to deviate from the MECP\@.\cite{thiophosgene,weird,carbene,housane}
In fact, in some cases, singularities in the golden-rule correlation function can push the hopping point off the crossing seam altogether.\cite{oxygen,GRperspective,noncrossing}
Our more general method covers all of these cases.

The high-bridge instanton is shown in \Fig{fig:inst_bridge_regimes}(d). 
Unlike in the low-bridge case, the high-bridge instanton only spends an infinitesimally small time on $\VB$.
Additionally, the energy of the trajectory on $\VB$ is not the same as the tunneling energy of the instanton---a consequence of the fact that the instanton is not a stationary point of $\tau'$ and $\tau''$.
Therefore, in the high-bridge regime, transitions to the virtual bridge states are accounted for by a nonresonant hop %
that
does not conserve energy.
However, the energy conservation is only violated for an infinitesimally short time and does not affect the overall energy of the reaction, 
as the energies of the trajectories on $\VR$ and $\VP$ are identical---a consequence of the stationary-point condition of $\tau$.

\subsection{A uniform approximation} \label{sec:bridge_uniform}

As the system transitions between the high- and low-bridge regimes,
the effective coupling term \eqn{eq:Delta_highbridge} will diverge. %
This is because, as depicted in \Fig{fig:bridge_cff}(c), the first-order asymptotic approximation of $\phi$ [\eqn{eq:phiTaylor}] overestimates the exact integral and diverges when the stationary point approaches the origin. %
Further, in the borderline low-bridge case [\Fig{fig:bridge_cff}(b)], the Gaussian approximation overestimates the exact integral %
by incorrectly including the
region $\tau', \tau'' < 0$.
A uniform asymptotic theory is therefore required to connect the high- and low-bridge regimes.

In this section, we will develop a uniform approximation that smoothly connects the low- and high-bridge regimes with different asymptotic expansions following the rules of asymptotic analysis.\cite{BenderBook}
This will chiefly be important for the crossover region where the system transitions between the low- and high-bridge regimes, 
as shown in \Figs{fig:inst_bridge_regimes}(b) and (c).
As the bridge rises, the segments of the instanton trajectory on the bridge state get shorter until the four hopping points coalesce. %
Here,
the asymptotic behavior of the correlation function changes,
leading to different instanton rate expressions for the high-bridge and low-bridge regimes.
In particular, the different asympototic behavior causes
$k\subtxt{LB}$ to overestimate the exact rate constant
and $k\subtxt{HB}$ to diverge, as can be inferred from \Figs{fig:bridge_cff}(b) and (c).
In order to describe the intermediate case within a uniform theory, we need a consistent asymptotic treatment which works in both regimes.

It will be necessary to perform the integral in two steps,
first integrating over
$\vec{u} = (\vec{x}_-', \vec{x}_-'', \vec{x}', \vec{x}'', z)$
and then over $\vec{v}=(\tau',\tau'')$.
The functions will be expanded around
$\tilde{\vec{u}}$ and $\tilde{\vec{v}}$, which are defined by the instanton geometry in the relevant regime.
Therefore, in the low-bridge regime, both $\tilde{\vec{u}}$ and $\tilde{\vec{v}}$ are stationary points while in the high-bridge regime, only  $\tilde{\vec{u}}$ is a stationary point, %
whereas $\tilde{\vec{v}}=\vec{0}$ is not.
Nonetheless, in both cases, the choice reflects the dominant point of the integrand over the integration domain.

First, expanding the action as a Taylor series in $\vec{u}$ around the stationary point, $\tilde{\vec{u}}\equiv\tilde{\vec{u}}(\vec{v})$, gives
\begin{align} \label{eq:S4_expansion_u}
	S^{(4)} & \simeq S^{(4)}(\tilde{\vec{u}}, \vec{v}) + \frac{1}{2} (\vec{u}-\tilde{\vec{u}})\T \mat{H}_{\vec{u}} (\vec{u}-\tilde{\vec{u}}) ,
\end{align}
with the $\vec{v}$-dependent matrix
\begin{align}
	\mat{H}_{\vec{u}} &= \pdv{S^{(4)}}{\vec{u}}{\vec{u}} \bigg|_{\vec{u}=\tilde{\vec{u}}} .
\end{align}
This allows us to integrate over $\vec{u}$ (assuming slowly varying prefactors) to obtain
\begin{align} \label{eq:k4U_int1}
	k \ZR \simeq \int_{0}^{\infty} \rmd \vec{v} \,  & \frac{|\DeltaRB\DeltaBP|^2}{\hbar^4} \sqrt{\frac{\CR\CB'\CP\CB''}{-\det[\mat{H}_{\vec{u}}]}} \nonumber \\ 
    & \times  \sqrt{2\pi\hbar} \; \eu{-\phi/\hbar} ,
\end{align}
with the effective action
\begin{align} \label{eq:phi}
	\phi(\vec{v}) = S^{(4)}(\tilde{\vec{u}}(\vec{v}), \vec{v}) .
\end{align}

Next, we expand $\phi$ as a second-order Taylor series in $\vec{v}$ around $\tilde{\vec{v}}$ to give
\begin{align}
	\phi(\vec{v})
	\simeq
	\phi(\tilde{\vec{v}})
	+ \vec{\mathcal{E}}\T(\vec{v}-\tilde{\vec{v}})
	+ \frac{1}{2} (\vec{v} - \tilde{\vec{v}})\T [ \mat{H}/\mat{H}_{\vec{u}} ] (\vec{v} - \tilde{\vec{v}}) ,
\end{align}
where $\vec{\mathcal{E}} = \frac{\partial{S^{(4)}}}{\partial{\vec{v}}}$,
and $[\mat{H}/\mat{H}_{\vec{u}}]$ is the Schur complement of the $\vec{u}$ block of $\mat{H}$,
\begin{align} \label{Schur}
	\mat{H}/\mat{H}_{\vec{u}} = \pdv{S^{(4)}}{\vec{v}}{\vec{v}} - \pdv{S^{(4)}}{\vec{v}}{\vec{u}} \left[ \pdv{S^{(4)}}{\vec{u}}{\vec{u}} \right]^{-1} \pdv{S^{(4)}}{\vec{u}}{\vec{v}} ,
\end{align}
evaluated at the instanton configuration $(\tilde{\vec{u}}(\tilde{\vec{v}}),\tilde{\vec{v}})$.

In the low-bridge case, $\vec{\mathcal{E}} = (\mathcal{E}',\mathcal{E}'')$ [as defined previously in Eqs.~\eqref{eq:mathcalE} and \eqref{eq:energydiff_MECP}] is $\vec{0}$ due to the stationary-point condition with respect to $\tau'$ and $\tau''$.
Using $\Sigma^{(4)} = \det[\mat{H}] = \det[\mat{H}/\mat{H}_{\vec{u}}] \det[\mat{H}_{\vec{u}}]$, %
we obtain
\begin{align} \label{eq:k4U}
		k\subtxt{U} = \mathcal{A}(\tilde{\vec{v}}) k\subtxt{LB} ,
\end{align}
where the integral
\begin{align} \label{eq:uniform_bridge_correction_factor_LB}
   \mathcal{A}(\tilde{\vec{v}}) =  \frac{\sqrt{\det[\mat{H}/\mat{H}_{\vec{u}}]}}{2\pi\hbar} \int_{0}^{\infty} \rmd \vec{v} \; \eu{-(\vec{v} - \tilde{\vec{v}})\T [ \mat{H}/\mat{H}_{\vec{u}} ]  (\vec{v} - \tilde{\vec{v}}) /2\hbar}
\end{align}
is a two-dimensional analogue of the error function and can be evaluated numerically.
It recovers the low-bridge rate expression [\eqn{eq:k4_inst}] in the limit that $\hbar\rightarrow0$ as long as $\tilde{\vec{v}}>\vec{0}$. %

The key difference from the derivation in \Sec{sec:low_bridge} %
is that we account for the fact that the lower bound of the integral  in \eqn{eq:k4U_int1} is 0 and not $-\infty$.
The factor $\mathcal{A}$, corrects for the overestimate of the integral that results from erroneously including the region $\tau',\tau'' < 0$.
The factor is always less than 1 and in certain limits, it approaches a value of 1/4.%
\footnote{\red{In particular, in symmetric reactions, the Schur complement takes a particularly simple form such that the integral in \eqn{eq:uniform_bridge_correction_factor_LB} decouples into two error functions, each of which evaluates to 1/2 in the limit that $\tilde{\vec{v}}\to\mathbf{0}$}}

In the high-bridge case,
where $\tilde{\vec{v}}=\vec{0}$,
the determinant decouples to leading order such that
$\det[\mat{H}_{\vec{u}}] \simeq C\subtxt{B}' C\subtxt{B}'' \Sigma^{(2)}$.
We can therefore tidy up the prefactors of \eqn{eq:k4U_int1} to obtain
\begin{align}
	k \ZR \simeq \int_{0}^{\infty} \rmd \vec{v} \,  & \frac{|\DeltaRB\DeltaBP|^2}{\hbar^4} \sqrt{\frac{\CR\CP}{-\Sigma^{(2)}}} \nonumber \\ 
    & \times  \sqrt{2\pi\hbar} \; \eu{-\phi/\hbar} .
\end{align}
Integrating over $\vec{v}$ gives %
an expression similar to \eqn{eq:kHB}:
\begin{align} \label{eq:k2U}
    k\subtxt{U} \ZR = \sqrt{2\pi\hbar} \, \frac{|\Delta\subtxt{SEC}\suptxt{U}|^2}{\hbar^2} \sqrt{\frac{\CR\CP}{-\Sigma^{(2)}}} \; \eu{-S^{(2)}/\hbar} ,
\end{align}
with a new effective coupling
\begin{align} \label{eq:rate_uniformbridge_int2}
	|\Delta\subtxt{SEC}\suptxt{U}| &= \frac{|\DeltaRB \DeltaBP|}{\hbar} 
    \sqrt{\mathcal{B}} ,
\end{align}
where
\begin{align}
    \mathcal{B} &= \int_{0}^{\infty} \rmd \vec{v} \, \eu{-(\vec{\mathcal{E}}\T \vec{v} + \frac{1}{2} \vec{v}\T [\mat{H}/\mat{H}_{\vec{u}}] \vec{v} )/\hbar} .
\end{align}
In the limit of $\hbar\rightarrow0$, the expression is dominated by the linear exponent and thus recovers the original high-bridge rate \eqn{eq:kHB}.
However, because of the inclusion of second-derivative terms, the expression no longer diverges when $\vec{\mathcal{E}} \to \vec{0}$, but instead connects smoothly with the uniform approximation in the low-bridge regime.

Note that, in order to evaluate %
$[\mat{H}/\mat{H}_{\vec{u}}]$
in the high-bridge regime,
all the required quantities can be obtained from the two-state golden-rule instanton between states R and P along with one additional evaluation of $\VB$, its gradient and Hessian at the hopping point.
The derivatives of the bridge action are described using the short-time limit expressions \eqn{eq:shorttime}, with $\tilde{\vec{x}}_-'$ and $\tilde{\vec{x}}_-''$ expressed in terms of the hopping-point momenta  $\vec{p}_{\mathrm{B}}'$ and $\vec{p}_{\mathrm{B}}''$, using \eqn{eq:pB}.
In particular,
\begin{subequations}
\label{eq:hessterms_uniform_highbridge}
\begin{align} 
    \pdv{S^{(4)}}{\vec{v}}{\vec{v}} &= \begin{pmatrix}
        \frac{||\vec{p}_{\mathrm{B}}'||^2}{m\tau'} & 0 \\
        0 & \frac{||\vec{p}_{\mathrm{B}}''||^2}{m\tau''}
    \end{pmatrix} + \pdv{(\SR + \SP)}{\vec{v}}{\vec{v}}, \\[0.5em]
    \pdv{S^{(4)}}{\vec{u}}{\vec{v}} &= \begin{pmatrix}
         \frac{\vec{p}\subtxt{B}'}{\tau'} & \mathbf{0} \\
         \mathbf{0} & -\frac{\vec{p}\subtxt{B}''}{\tau''} \\
         \pdv{V\subtxt{B}}{\vec{x}'} & \mathbf{0} \\
         \mathbf{0} & \pdv{V\subtxt{B}}{\vec{x}''} \\
         0 & 0
     \end{pmatrix} + \pdv{(\SR + \SP)}{\vec{u}}{\vec{v}} , \\[0.5em]
    \pdv{S^{(4)}}{\vec{u}}{\vec{u}} &= \diag\left(\frac{m}{\tau'}\mathbb{I}_{f}, \frac{m}{\tau''}\mathbb{I}_{f}, \tau'\pdv{V\subtxt{B}}{\vec{x}'}{\vec{x}'}, \tau''\pdv{V\subtxt{B}}{\vec{x}''}{\vec{x}''}, 0\right) \nonumber\\
    &\quad + \pdv{(\SR + \SP)}{\vec{u}}{\vec{u}}
\end{align}
\end{subequations}

As a result of the $1/\tau'$ and $1/\tau''$ terms, $\pdv{S^{(4)}}{\vec{v}}{\vec{v}}$, $\pdv{S^{(4)}}{\vec{u}}{\vec{v}}$, %
and $\pdv{S^{(4)}}{\vec{u}}{\vec{u}}$ individually diverge.
However, the divergences cancel exactly in the final expression \eqn{Schur}, leaving a well-defined remainder.\footnote{For example, consider the first entry of the matrix:
\begin{align*}
    &\pdv{S^{(4)}}{\tau'}{\tau'} - \pdv{S^{(4)}}{\tau'}{\vec{u}}\left(\pdv{S^{(4)}}{\vec{u}}{\vec{u}}\right)^{-1}\pdv{S^{(4)}}{\vec{u}}{\tau'} = \left(\frac{||\vec{p}\subtxt{B}'||^2}{m\tau'} + \mathcal{O}(\tau')^0\right)  \\
    & \quad - \left(\frac{\vec{p}\subtxt{B}'}{\tau'} + \mathcal{O}(\tau')^0\right)\cdot\left(\frac{m}{\tau'} \mathbb{I}_f + \mathcal{O}(\tau')^0\right)^{-1}\cdot\left(\frac{\vec{p}\subtxt{B}'}{\tau'} + \mathcal{O}(\tau')^0\right) \\ %
    &= \mathcal{O}(\tau')^0,
\end{align*}
which has no divergence as $\tau'\to0$.}
However, explicitly writing down a formula for this remainder involves lengthy algebraic expressions.
For this reason, it is easiest to evaluate the Schur complement \eqn{Schur}
using \eqn{eq:hessterms_uniform_highbridge} with
small but finite values for $\tau'$ and $\tau''$. %

\section{Results and discussion} \label{sec:results}

We now benchmark the newly derived instanton methods against quantum-mechanical results for the model system given by \eqn{eq:model_system}.
The quantum-mechanical rate constants are calculated using \eqn{eq:rate_bridge}, with analytical formulas used for the wavefunction overlaps.\cite{harmonic_overlaps}
The Dirac delta function $\delta(\ER-\EP)$ is replaced by the sinc function $\sin(\pi(\ER - \EP)/\hbar\omega)/\pi(\ER - \EP)$, which allows one to formally define a rate constant for the one-dimensional model system that would otherwise exhibit coherent dynamics between reactant and product energy levels.
This is equivalent to truncating the integration range of the correlation function over the $t$ variable to remove the recurrences (which would anyway be washed out in a more realistic multidimensional system). %
As all the PESs have equal frequencies and the bias of $\VP$ was chosen to be an integer multiple of $\hbar\omega$,
the sinc function evaluates to $1/\hbar\omega$ when $\ER = \EP$ and is zero otherwise.

All the results correspond to a temperature of \SI{50}{\kelvin} and constant couplings of
$|\DeltaRB| = \SI{40}{\per\cm}$ and $|\DeltaBP| = \SI{100}{\per\cm}$, 
with the remaining parameters as described in \Sec{sec:low_bridge}.
The system qualitatively
corresponds to the high-spin (quintet) $\to$ low-spin (singlet) decay of an iron(II) polypyridyl complex via a triplet bridge.\cite{liesst}
For the uniform approximation in the high-bridge regime, $\tau'=\tau''=0.001$ (in atomic units of time) was used to evaluate the Hessian terms in \eqn{eq:hessterms_uniform_highbridge}.
This was found to be numerically stable and give results converged to the number of significant figures reported.

\begin{figure}
	\centering
	\includegraphics{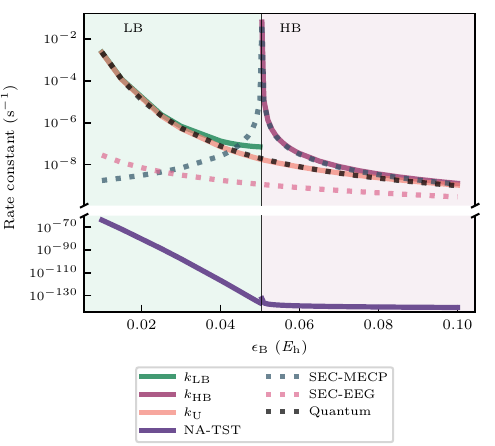}
	\caption{Instanton ($k\subtxt{HB}$, $k\subtxt{LB}$ and $k\subtxt{U}$), superexchange effective coupling (SEC) and nonadiabatic transition-state theory (NA-TST) rate constants benchmarked against the quantum-mechanical rate constants as a function of the bridge bias. 
    The quantum-mechanical and SEC rate constants were calculated from Frank--Condon overlaps of harmonic oscillator wavefunctions.
    SEC rate constants are calculated using Fermi's golden rule with two effective coupling terms.
    SEC-MECP  uses the potential energy difference at the reactant--product MECP [\eqns{eq:Delta_highbridge}], while SEC-EEG uses the electronic energy gap [\eqn{eq:Delta_highbridge_SEC}].
    NA-TST rate constants are given by \eqn{eq:k_NATST_low-bridge_Ulstrup} for the low-bridge regime and \eqn{eq:k_NATST_high-bridge} for the high-bridge regime.
	The background colors distinguish the two regimes of the PESs. }
	\label{fig:rates_model_system_Fepic3}
\end{figure} 

\subsection{Nonadiabatic transition-state theory}

For comparison, we additionally consider classical transition-state theory, which does not include tunneling contributions.
For the high-bridge regime, we use golden-rule NA-TST \cite{Lorquet1988NATST,Harvey1999NATST,GoldenGreens} with an effective coupling.\cite{UlstrupBook} 
Here, the reactant--product MECP at $\xRP$ plays the role of the transition state with energy $V\subtxt{RP}$, leading to the rate constant
\begin{align} \label{eq:k_NATST_high-bridge} 
	k\subtxt{HB}\suptxt{NA-TST} \ZR = \frac{|\Delta\subtxt{SEC}\suptxt{MECP}|^2}{\hbar^2 ||\vec{g}\subtxt{P} - \vec{g}\subtxt{R}||} \sqrt{\frac{2\pi m}{\beta}} \; \eu{-\beta V\subtxt{RP}} ,
\end{align}
where $\vec{g}_n$ is the gradient of $V_n$ at $\xRP$ and $\Delta\subtxt{SEC}\suptxt{MECP}$ is defined equivalently to \eqn{eq:Delta_highbridge} but using quantities evaluated at the MECP.

For the low-bridge regime, the higher energy of the two MECPs between $\VR$ and $\VB$ (at energy $\VRB$) and between $\VB$ and $\VP$ (at energy $\VBP$) plays the role of the transition state with energy $\VTS$. The rate constant then reads \cite{UlstrupBook} 
\begin{align}  \label{eq:k_NATST_low-bridge_Ulstrup}
	k\subtxt{LB}\suptxt{NA-TST} \ZR &=
    \frac{|\DeltaRB|^2}{\hbar^2 ||\vec{g}\subtxt{R(B)} - \vec{g}\subtxt{B(R)}||} \frac{|\DeltaBP|^2}{\hbar^2||\vec{g}\subtxt{B(P)} - \vec{g}\subtxt{P(B)}||} \nonumber \\
	& \quad \times  \frac{4m\hbar\pi^{3/2}  }{ \sqrt{\beta|\VRB - \VBP|} } \; \eu{-\beta\VTS}  ,
\end{align}
where $\vec{g}_{m(n)}$ is the gradient of $V_m$ at the MECP between $V_m$ and $V_n$.\footnote{Note that compared to the original expression (Eq. 7.20 in \Refx{UlstrupBook}), \eqn{eq:k_NATST_low-bridge_Ulstrup} differs by a factor of $2/\beta$.
The factor of 2 was added to account for recrossing events (see Eq. 7.33 in \Refx{UlstrupBook}), while the $\beta$ factor corrects an erroneous $\sqrt{\beta}$ in the numerator instead of the denominator.}

Classical theories neglect nuclear quantum effects such as tunneling and consequently significantly underestimate the
quantum-mechanical benchmark in \Fig{fig:rates_model_system_Fepic3} by orders of magnitude.
This is particularly pronounced due to the low temperature, where tunneling effects %
are particularly important.
Note that simple one-dimensional tunneling corrections to transition-state theory such as the weak-coupling method (which is based on the assumption of linear potentials)\cite{Lykhin2016NATST} result in an inconsistent description of the PESs and are particularly unreliable at low temperatures.\cite{thiophosgene,nitrene,housane}
It is for these reasons that a method is required to capture tunneling effects without such ad hoc assumptions.

Note that the two NA-TST results do not smoothly connect between the low-bridge and high-bridge regimes, and in fact the high-bridge result diverges at the crossover.
In this sense, the behavior is similar to the instanton theories, which was fixed using a uniform asymptotic approximation.
It may therefore also be useful to develop a uniform asymptotic approximation for the high-temperature limit where NA-TST is valid. 
However, this would require a separate investigation; one cannot simply take the high-temperature limit of the semiclassical instanton methods presented in this work, as they are only valid when the temperature is sufficiently low that there is no significant thermal population of the bridge states.\cite{PhilTransA}

\subsection{Superexchange Effective Coupling}

The SEC rate constants are calculated by using the usual thermal FGR expression in terms of vibrational wavefunction overlaps,\cite{DiracBook} with two variants for the energy-difference term of the effective coupling.
\red{The SEC-EEG method uses the electronic energy gaps between} the bridge and reactant states [\eqn{eq:Delta_highbridge_SEC}] while the SEC-MECP uses the potential energy difference at the reactant--product MECP [equivalent to \eqn{eq:Delta_highbridge}]. %

Note that as SEC uses vibrational wavefunction overlaps (also known as Franck--Condon factors), it inherently includes nuclear quantum effects such as tunneling and zero-point energy.
As a result, both variants perform significantly better than classical transition state theory, as shown in \Fig{fig:rates_model_system_Fepic3}.

\red{However, %
SEC-EEG still has significant errors, and although they appear to decrease at larger biases, it is not a valid $\hbar\to0$ asymptotic approximation to the quantum-mechanical benchmark in either regime.
This is because, as we have demonstrated in \Sec{sec:high_bridge}, the correct energy-difference term in the high-bridge regime corresponds to the one used by SEC-MECP, 
which is in good agreement with the quantum-mechanical benchmark, especially as the system is pushed further into the high-bridge regime.}

On the other hand, the SEC-MECP theory diverges in the crossover region, where the potential energy difference approaches 0.
Additionally, using SEC-MECP for the low-bridge regime gives completely incorrect behavior.
This is also explained by our instanton analysis, which indicates that the SEC approximation is only valid in the high-bridge regime, whereas in the low-bridge regime, one must explicitly account for tunneling dynamics on the bridge.

\subsection{Instanton Theory}

As expected from a rigorous asymptotic approximation, 
we find that the low-bridge and high-bridge instanton methods show excellent agreement with the quantum-mechanical benchmark in their respective regimes of applicability.
In the high-bridge regime,
the instanton rate is the semiclassical approximation to the SEC-MECP approach and thus follows it closely.
However, in the low-bridge regime, the instanton method overcomes the limitations of the SEC approach, which is not formally valid.

In the crossover region where the system transitions between the low- and high-bridge regimes, the leading-order asymptotic approximations break down.
Like SEC-MECP, the high-bridge instanton method diverges, due to the energy-difference term of the effective coupling [\eqn{eq:Delta_highbridge}] approaching 0, while the low-bridge instanton method overestimates the rate constant by about a factor of 4, as a result of including regions outside the integration bounds of $\tau'$ and $\tau''$ [see \Fig{fig:bridge_cff}(b)]. 
Both these problems are fixed by the uniform instanton theory. %
As expected of a uniform approximation, it tends to $k\subtxt{HB}$ and $k\subtxt{LB}$ in the appropriate limits but avoids the divergence and smoothly connects the two expressions at the boundary between the two regimes.

Overall, we find that the instanton methods are more accurate across a wide range of cases.
This contrasts with SEC methods, which are only valid for high bridges and, even after reducing the three-state nonadiabatic problem into an effective two-state problem, still require the evaluation of vibrational wavefunctions and their overlaps, which are intractable for anything but simplified model systems.
On the other hand, instanton theory is more computationally efficient when applied to real molecular systems without making global harmonic approximations to the PESs.

\section{Conclusions} \label{sec:conclusions}

We have extended nonadiabatic instanton theory to account for bridge-mediated processes in the high-bridge regime,
where we find that the three-state bridge-mediated instanton reduces to a two-state golden-rule instanton with an effective coupling.
Unlike the low-bridge regime, where imaginary-time trajectories on the bridge PES account for virtual transitions to bridge states, in the high-bridge regime, infinitesimally short imaginary-time trajectories describe a nonresonant hop. 
The resulting theory is similar to the well-known SEC expression, 
except that our rigorous derivation provides the correct effective coupling that
obeys detailed balance
and accounts for corner cutting in multidimensional systems.
We have also developed a uniform instanton approximation, which smoothly connects the instanton rate expressions in the low-bridge and high-bridge regimes with the correct asymptotic behavior.
We find that the resulting uniform instanton theory is accurate across a wide range of cases. %

In the low-bridge regime, the instanton trajectory consists of four segments: one on the reactant, one on the product and two on the bridge.
On the other hand, in the 
high-bridge regime it consists of two segments: one on each of the reactant and product PESs.
It is therefore crucial to identify the relevant regime in order to apply the appropriate instanton theory. %
The recommended workflow for calculating instanton rate constants for bridge-mediated processes is as follows:
first, the reactant--product MECP is located and $\VR$, $\VP$ and $\VB$ are calculated at this point.
The relevant regime is identified based on whether the energy of $\VB$ is above or below $\VR$ and $\VP$, and the appropriate instanton pathway is optimized.
Note that as the instanton hopping points are not necessarily at the MECP, the potentials along the instanton pathway should be double-checked to ensure that the correct regime was chosen.
In a borderline case, the uniform approximation must be used.
This is especially important in the high-bridge regime, where the leading-order instanton rate can diverge.
However, as the maximum extra work required to evaluate the uniform correction is one additional gradient and Hessian calculation on the bridge at the hopping points, %
it is recommended to be used in general.

With the extension to the high-bridge regime,
we significantly expand the applicability of bridge-mediated instanton theory.
In a companion work,\cite{liesst} we demonstrate the applicability of the new method in conjunction with on-the-fly electronic-structure calculations to study light-induced excited spin-state trapping (LIESST) phenomena, where an iron(II) complex is trapped in a metastable quintet state following a photoexcitation of the ground singlet state.\cite{Fepic3_liesst_OG,Hauser2004,Fepic3_PESs}
The quintet $\to$ singlet decay that follows is a bridge-mediated nonadiabatic process via the triplet state. 

However, bridge-mediated instanton theory is not restricted to LIESST complexes but is applicable to any nonadiabatic bridge-mediated process, such as double intersystem crossings\cite{liesst_review_2,Hauser2004,Feterpy2_rocking,lundberg_transitionmetal_dimers} or through-bond electron transfers,\cite{Franzen1993SupexET,quantumbiology_review_1,Bixon1989_PhotosynthesysSuperexchange,Jortner1995BacRCSuperexET} as long as the relevant diabatic couplings are small.
For even broader applicability, such as intersystem crossings of second and third-row transition metal complexes, where spin--orbit couplings are typically larger, higher-order correction terms to the bridge-mediated instanton could be derived, analogous to a similar extension of golden-rule instanton theory.\cite{4thorder} 
One could also further extend bridge-mediated instanton theory to account for multiple bridge states, which would be useful for reactions with long-range electron transfer with multiple bridge states in biomolecules,\cite{giese_multistep_ET,giese_longrange_ET_DNA,jiaxing_twoelectron_ET} two-electron electron transfers,\cite{jiaxing_twoelectron_ET} long-range Dexter energy transfer,\cite{bai_dexter,skourtis2016dexter} and for systems with three or more intersystem crossings.\cite{Pt_cluster_tunnelling,benzene_chromium_dissociation}

\section*{Supplementary Material}
See the supplementary material for the numerical data plotted in Fig.~\ref{fig:rates_model_system_Fepic3}.

\section*{Acknowledgements}
The authors acknowledge financial support from the SNSF project titled ``Nonadiabatic effects in chemical reactions'' (Grant Number - 207772).

\section*{AUTHOR DECLARATIONS}
\subsection*{Conflict of Interest}
The authors have no conflicts to disclose.

\subsection*{Author Contributions}
\textbf{Imaad M. Ansari}: Conceptualization (equal); Formal Analysis (equal); Methodology (equal); Investigation (lead); Software (lead); Visualization (lead); Writing – original draft (lead);  Writing – review \& editing (supporting).
\textbf{George Trenins}: Conceptualization (equal); Formal analysis (supporting); Methodology (supporting); Supervision (supporting); Writing - review \& editing (supporting).
\textbf{Jeremy O. Richardson}: Conceptualization (equal); Formal analysis (equal); Funding acquisition (lead); Methodology (equal); Supervision (lead); Validation (lead); Writing – original draft (supporting); Writing – review \& editing (lead).

\section*{DATA AVAILABILITY}
The data that supports the findings of this study are available within the article and its
supplementary material.

\bibliography{references, bridge}

\end{document}


\title{Supplementary information for:\\\PaperTitle}

\author{Imaad M. Ansari, George Trenins and Jeremy O. Richardson}

\maketitle

\FloatBarrier

\begin{table}[h]
\centering
\caption{Instanton and transition-state theory rate constants in the low-bridge regime}
\sisetup{table-number-alignment = center, round-mode=places, round-precision=3, exponent-mode = scientific}
\begin{tabular}{@{}S[round-precision=4, exponent-mode = input]SSS@{}}
\toprule
 \head{Bridge energy (\SI{}{\electronvolt})} & \head{$k\subtxt{LB}$ (\SI{}{\per\second}) } & \head{$k\subtxt{U}$ (\SI{}{\per\second}) } & \head{$k\subtxt{NA-TST}$ (\SI{}{\per\second}) } \\ \midrule
0.01  &  0.002322872124427378    & 0.002254887607505238   & 1.8526761779219067e-64  \\
0.015 &  0.00012149130974234887  & 0.0001160048663962857  & 1.7524547585228726e-72  \\
0.025 &  2.443376699734423e-06   & 2.12430055107477e-06   & 1.7022053525337586e-89  \\
0.03  &  6.757101643345069e-07   & 5.319692290309255e-07  & 1.7632939835226033e-98  \\
0.04  &  1.327425030756058e-07   & 7.16040099957648e-08   & 2.2239967104457093e-117 \\
0.041 &  1.1956340546405845e-07  & 6.105717084659247e-08  & 2.468917912230536e-119  \\
0.042 &  1.0871301149546666e-07  & 5.236893391178745e-08  & 2.6743996794278563e-121 \\
0.043 &  9.975916666217535e-08   & 4.516379908338644e-08  & 2.8307270186002642e-123 \\
0.044 &  9.236638435605048e-08   & 3.915065654557314e-08  & 2.9335397530817504e-125 \\
0.045 &  8.627233606036463e-08   & 3.410185157711015e-08  & 2.9855054247243196e-127 \\
0.047 &  7.72004107461604e-08    & 2.6217142299082666e-08 & 2.9970480272048086e-131 \\
0.048 &  7.393383987165665e-08   & 2.3125904421320846e-08 & 3.0338340309709596e-133 \\
0.049 &  7.137197594622262e-08   & 2.0473411016361374e-08 & 3.253082583827585e-135  \\
0.05  &  6.943826364647843e-08   & 1.8186358382794915e-08 & 4.557783465870336e-137  \\
0.0504 &  6.88335e-08 & 1.73617e-08  & 1.51126e-137 \\
\bottomrule
\end{tabular}
\end{table}

\begin{table}[h]
\centering
\caption{Instanton and transition-state theory rate constants in the high-bridge regime}
\sisetup{table-number-alignment = center, round-mode=places, round-precision=3, exponent-mode = scientific}
\begin{tabular}{@{}S[round-precision=4, exponent-mode = input]SSS@{}}
\toprule
 \head{Bridge energy (\SI{}{\electronvolt})} & \head{$k\subtxt{HB}$ (\SI{}{\per\second}) } & \head{$k\subtxt{U}$ (\SI{}{\per\second}) } & \head{$k\subtxt{NA-TST}$ (\SI{}{\per\second}) } \\ \midrule
0.0505 & 0.06501228676395925    & 1.7125509510409365e-08 & 1.5483466050045717e-133 \\
0.051  & 1.1892070862158003e-05 & 1.633354507499439e-08  & 2.8322411750795e-137 \\
0.052  & 1.345495235781522e-06  & 1.4886051956857025e-08 & 3.2044603936729354e-138 \\
0.053  & 4.861459692082051e-07  & 1.3600233152217463e-08 & 1.1578156967360622e-138 \\
0.054  & 2.484216896285227e-07  & 1.2454980869505185e-08 & 5.91646439299004e-139 \\
0.055  & 1.5041043416250787e-07 & 1.1432285570851973e-08 & 3.5822072516589e-139 \\
0.056  & 1.0074369084833062e-07 & 1.0516736691093686e-08 & 2.3993334101134356e-139 \\
0.057  & 7.215773379696175e-08  & 9.695110286844664e-09  & 1.718524108450316e-139 \\
0.058  & 5.42137052966999e-08   & 8.956027450184775e-09  & 1.2911652661231758e-139 \\
0.059  & 4.2216972556442404e-08 & 8.289670504611884e-09  & 1.0054484988147795e-139 \\
0.06   & 3.380270131286737e-08  & 7.68754652467769e-09   & 8.050524050597845e-140 \\
0.065  & 1.4517057978127962e-08 & 5.413501252153821e-09  & 3.4574137526800156e-140 \\
0.07   & 8.028800043111458e-09  & 3.962044830900199e-09  & 1.9121562873410176e-140 \\
0.075  & 5.086862423981832e-09  & 2.9951548759167773e-09 & 1.211498096182017e-140 \\
0.08   & 3.508967212279164e-09  & 2.3269165411296867e-09 & 8.357031786823318e-141 \\
0.085  & 2.565750356200287e-09  & 1.8501688960486609e-09 & 6.110646234819583e-141 \\
0.09   & 1.9574029533593277e-09 & 1.5005324306169653e-09 & 4.661792975324141e-141 \\
0.095  & 1.5423088228228365e-09 & 1.2378840325886593e-09 & 3.673195865816221e-141 \\
0.1    & 1.2465063081489972e-09 & 1.0363893279456176e-09 & 2.9687062344793973e-141 \\
\bottomrule
\end{tabular}
\end{table}

\begin{table}[h]
\centering
\caption{Rate constants evaluated using nuclear wavefunction overlaps}
\sisetup{table-number-alignment = center, round-mode=places, round-precision=3, exponent-mode = scientific}
\begin{tabular}{@{}S[round-precision=4, exponent-mode = input]SSS@{}}
\toprule
 \head{Bridge energy (\SI{}{\electronvolt})} & \head{Quantum (\SI{}{\per\second}) } & \head{ SEC-MECP (\SI{}{\per\second}) } & \head{ SEC-ZPE (\SI{}{\per\second}) } \\ \midrule
0.01   & 0.002341463457862328   & 1.7181362338781052e-09 & 2.8172190183703285e-08 \\
0.015  & 0.00011817939523772019 & 2.236308156465109e-09  & 1.2520973414979232e-08 \\
0.02   & 1.2484720762642008e-05 & 3.0298154582737032e-09 & 7.043047545925821e-09 \\
0.025  & 2.1863516916786506e-06 & 4.334847096813792e-09  & 4.507550429392524e-09 \\
0.03   & 5.539939961224096e-07  & 6.708159659894202e-09  & 3.130243353744809e-09 \\
0.04   & 7.633207406015203e-08  & 2.5586405167404126e-08 & 1.7607618864814553e-09 \\
0.041  & 6.525926513129085e-08  & 3.126082198858337e-08  & 1.6759185118205399e-09 \\
0.042  & 5.612920008470472e-08  & 3.905562086315608e-08  & 1.597062935584086e-09 \\
0.043  & 4.8552426544440234e-08 & 5.017557425440773e-08  & 1.5236446827313838e-09 \\
0.044  & 4.222601182603358e-08  & 6.682060824257856e-08  & 1.4551751127945913e-09 \\
0.045  & 3.691264724014344e-08  & 9.336379662769116e-08  & 1.3912192683310264e-09 \\
0.047  & 2.8614987879667814e-08 & 2.3088058467684824e-07 & 1.2753368122998317e-09 \\
0.048  & 2.5363422058116716e-08 & 4.532372304483277e-07  & 1.2227513100565661e-09 \\
0.049  & 2.2575089577225442e-08 & 1.2636206394218002e-06 & 1.1733523608372876e-09 \\
0.0505 & 1.9096555992220197e-08 & 0.05995023574215112    & 1.1046834696089906e-09 \\
0.051  & 1.8094208140665067e-08 & 1.096611866980144e-05  & 1.0831291881469933e-09 \\
0.052  & 1.628782326688916e-08  & 1.2407309539488512e-06 & 1.0418709387464234e-09 \\
0.053  & 1.4711714218686198e-08 & 4.4829319056171485e-07 & 1.0029259588360015e-09 \\
0.054  & 1.3331195379954204e-08 & 2.2907883414046595e-07 & 9.661244918965462e-10 \\
0.055  & 1.2117509142306341e-08 & 1.3869902806003842e-07 & 9.313120721885382e-10 \\
0.056  & 1.104670467388279e-08  & 9.289948587441445e-08  & 8.983479012660486e-10 \\
0.057  & 1.0098747449620378e-08 & 6.65393169056371e-08   & 8.671034220899748e-10 \\
0.058  & 9.256808492414957e-09  & 4.999246411363631e-08  & 8.374610637248301e-10 \\
0.059  & 8.506694302723035e-09  & 3.8929832852483304e-08 & 8.093131336886896e-10 \\
0.06   & 7.83638748290814e-09   & 3.117072192500289e-08  & 7.825608384362026e-10 \\
0.065  & 5.3721912695483705e-09 & 1.3386716440710006e-08 & 6.667974007977108e-10 \\
0.07   & 3.858855023170375e-09  & 7.403653667170469e-09  & 5.74942656810271e-10 \\
0.075  & 2.878673046099983e-09  & 4.690784106899935e-09  & 5.008389365991695e-10 \\
0.08   & 2.215144659844327e-09  & 3.2357485339868637e-09 & 4.401904716203638e-10 \\
0.085  & 1.7490286317076188e-09 & 2.365973362361203e-09  & 3.899265077329174e-10 \\
0.09   & 1.4111656064138516e-09 & 1.8049937071484135e-09 & 3.478048170827566e-10 \\
0.095  & 1.1596224954934426e-09 & 1.4222200466679555e-09 & 3.121572319523909e-10 \\
0.1    & 9.679783889239504e-10  & 1.149449600179852e-09  & 2.8172190183703276e-10 \\
\bottomrule
\end{tabular}
\end{table}